\documentclass[letterpaper,11pt]{article}
\usepackage{latexsym}
\usepackage{amssymb}
\usepackage{a4wide}
\begin{document}

\newtheorem{thm}{Theorem}
\newtheorem{lem}{Lemma}
\newtheorem{defn}{Definition}
\newtheorem{cor}{Corollary}

\newenvironment{Proof}{\begin{trivlist}
	\item[\hskip\labelsep{\bfseries Proof:}]}{\hfill$\Box$\end{trivlist}}
\newenvironment{claim}{\begin{trivlist}
	\item[\hskip\labelsep{\bfseries Claim:}]\itshape}{\end{trivlist}}

\def\CP{{\it Common Prefix }}
\def\G{\Gamma}
\def\S{\Sigma}
\def\E{OPT_{EBCS}}
\def\e{EBCS'}
\def\O{OPT_{CP}}
\def\N{OPT_{NN}}
\def\NN{{\it Nested Neighborhoods }}
\def\c{CP'}
\def\A{{\sf Algo-CP}}
\def\e{\epsilon}
\newcommand{\eat}[1]{}

\title{ The Common Prefix Problem On Trees }

\author{%
\begin{tabular}{p{5cm}p{5cm}}
  \centering Sreyash Kenkre & Sundar Vishwanathan \\
  \multicolumn{2}{c}{\small Department Of Computer Science \& Engineering,} \\
  \multicolumn{2}{c}{\small IIT Bombay, Powai-400076, India.}\\
  \multicolumn{2}{c}{\small {\tt \{srek,sundar\}@cse.iitb.ac.in}}
\end{tabular}%
}

\date{}
\maketitle

\begin{abstract}
We present a theoretical study of a problem arising in database query optimization~\cite{ravi}, which we
call as The \CP Problem. We present a $(1-o(1))$ factor approximation algorithm for this problem, when
the underlying graph is a binary tree. We then use a result of Feige and Kogan~\cite{Feige} to show
that even on stars, the problem is hard to approximate.
\end{abstract}

\section{Problem}
Let $T$ be a tree with $V$ as its vertex set and $E$ as its edge set. Let each vertex $v$ be associated with
a set of labels $S_v$, taken from an alphabet $\Sigma$. Suppose that the vertices $v$ and $u$ are adjacent and their
corresponding labels are given permutations $P_v$ and $P_u$. We define the {\it benefit} of the edge $uv$ as the length of 
the largest common prefix, denoted by $P_v\wedge P_u$. The goal is to maximize the total benefit by permuting the
labels associated with each vertex appropriately. More precisely, find permutations $P_1, P_2,\ldots, P_{|V|}$, so as
to maximize $\sum_{uv\in E} |P_u\wedge P_v|$. The corresponding decision problem is known to be $NP-Complete$~\cite{ravi}.
It can be solved in polynomial time if the tree is a path, and a $1/2$-factor approximation is known for the case
of a binary tree~\cite{ravi}. In this paper give a $(1-o(1))$ factor algorithm
for this problem on binary trees. We then study the problem when the underlying graph is a star ($K_{1,r}$)
and prove a hardness of approximation result by relating this problem to the {\it Maximum Edge Biclique} problem.
Throughout the paper we assume that the size of the alphabet $\S$ is a constant.

\section{Optimal Recursion For Trees}
In this section we give a recursion to optimally solve \CP on trees. This recursion may run in exponential time. In
the next section we will run this on sufficiently small trees to get the $(1-o(1))$ factor algorithm.


We observe that the labels that are common to all vertices can always be put as prefixes to the permutations associated
with the vertices. If the first label in the permutation associated with each vertex is the same, then we have
a label common to all vertices. Hence, once the common labels are removed, there will be an edge with zero benefit in
the optimal. This we can delete from the tree $T$, and recurse as follows.

\begin{eqnarray}\label{OPTrecursion}
\O(T) & = &  |\cap_{v\in V} S_v| + \max_{e\in E} [ \O(T_1) + \O(T_2) ]
\end{eqnarray}

\noindent where $T_1$ and $T_2$ are the two connected components of $T\setminus e$. However solving this
recursion may involve steps exponential in the number of nodes for example, on a complete binary tree of size $n$. 
The recursion-(\ref{OPTrecursion}) can be implemented as a dynamic program for trees which have a polynomially
bounded number of subtrees, for example, paths. We show that binary trees of height $\log\log n$ also have this
property.

\begin{claim}
The total number of subtrees in a binary tree of height $\log\log n$ is at most $n^2$.
\end{claim}
\begin{Proof}
The total number of nodes in a binary tree of height $h$ is at most $2^h$. Connecting each subset
of the vertices to the root yields a subtree containing the root, so there are at most $2^{2^h}$ such subtrees.
Thus the total number of subtrees in a binary tree of height $h$ is at most
\begin{eqnarray*}
 &      & 2^{2^h} + 2 2^{2^{h-1}} + 2^2 2^{2^{h-2}} + \ldots + 2^h \\
 &   =  & 2^{2^{h+1}} -2^h \\
 & \leq & 2^{2^{h+1}}
\end{eqnarray*}
\noindent If $h$ equals $\log\log n$, we get the desired result.
\end{Proof}

\noindent It follows that the recursion-(\ref{OPTrecursion}) can be solved optimally in time $O(n^2)$ on binary trees of height
$\log\log n$. We use this to give a $(1-\frac{1}{\log\log n})$ factor approximation for \CP on binary trees.

\section{$(1-O(1))$ Factor Algorithm}
Consider a binary tree $T$, of height $h$ on $n$ vertices rooted at vertex $r$. We split $T$ into sets $A_1,A_2,\ldots, A_{\log\log n}$,
each consisting of subtrees of height at most $\log\log n$. 
$A_1$ consists of the subtrees obtained by deleting the edges
joining vertices from heights $i\log\log n-1$ and $i\log\log n$ for $1\leq i \leq \lceil \frac{h}{\log\log n}\rceil$. 
$A_2$ consists of subtrees obtained by deleting the edges joining vertices from heights 
$i\log\log n$ and $i\log\log n +1$ for $0\leq i \leq \lceil \frac{h}{\log\log n}\rceil$ and so on.
Each $A_i$ consists of vertex disjoint subtrees of height {\it at most} $\log\log n$. Since each $A_i$ contains
no more than $n$ subtrees, we can solve \CP on each $A_i$ optimally. We denote the optimal value for $A_i$ by $\O(A_i)$.
Note that each edge occurs in all but one of the $A_i$'s. Let $b_e$ denote the benefit of the edge $e$ in the optimal,
and let $A$ denote the maximum of all $\O(A_i)$'s. Then from the preceding discussion we have,
\begin{eqnarray*}
(\log\log n -1)\O &   =  & (\log\log n -1)\sum_{e\in E} b_e \\
                  &   =  & \sum_{e\in A_1}b_e + \sum_{e\in A_2}b_e+ \ldots + \sum_{e\in A_{\log\log n}}b_e \\
	        & \leq & \O(A_1) + \O(A_2) + \ldots + \O(A_{\log\log n}) \\
	        & \leq & (\log\log n) A.
\end{eqnarray*}

We thus have a factor $(1-\frac{1}{\log\log n})$ algorithm for binary trees by taking the maximum of the $A_i$'s.
Since a binary tree of height $\log\log n$ has at most $n^2$ subtrees, and each $A_i$ can have at most 
$\frac{h}{\log\log n}$ trees, and since there are $\log\log n$ $A_i$'s, the total time taken for this algorithm is
$O(\frac{h}{\log\log n} n^2 \log\log n) = O(n^3)$. 

Note that we can trade the approximation factor for running time as follows. For fixed $\e<1$,
take $N=\lceil\frac{1}{\e}\rceil$. Now, instead of taking subtrees of height at most $\log\log n$ in the $A_i$'s
take them to be of height at most $N$. We can use the recursion-(\ref{OPTrecursion}) to solve for the subtrees of 
height at most $N$ in time $O(n2^{2^N})$. Using the same analysis as above, we get a $(1-\e)$ factor algorithm
that runs in $O(\frac{n}{\e}2^{2^{\lceil\frac{1}{\e}\rceil +1}})$ time.

\section{Common Prefix on Stars}
In this section we prove that the \CP problem on stars is equivalent to a problem of finding large nested
neighborhoods in bipartite graphs. We shall use this in the next section to prove a hardness of approximation result
for \CP. Consider the following problem.

\begin{defn} {\bf Nested Neighborhoods :}
Given a bipartite graph $G=(U,V,E)$ with $U$ and $V$ as its bipartition and $E$ as its edge set, find subsets
$U' \subseteq U$ and $V'\subseteq V$, such that the elements of $U'$ can be ordered as $u_1,u_2,\ldots,u_{|U'|}$, with
$\G(u_1)\cap V' \supseteq \G(u_2)\cap V' \supseteq \ldots \supseteq \G(u_{|U'|})\cap V'$, and such that
$|\G(u_1)\cap V'|+|\G(u_2)\cap V'|+\ldots+|\G(u_{|U'|})\cap V'|$ is maximized.
\end{defn}

\noindent Note that the above problem is independent of whether we choose the subset from $U$ or from $V$, since $V'$
can be labeled to get a feasible solution of the same cost. We show that this problem is equivalent to the \CP problem
on stars.

Suppose $G=(U,V,E)$ is an instance of {\it Nested Neighborhoods}. Consider a star $T$ with leaf nodes corresponding to
the vertices in $U$ and a vertex $r\not\in U$ as the non-leaf vertex. We treat the vertex set $V$ as a set of labels
to be assigned
to vertices of $T$. The vertex $r$ is given the entire set $V$ as its set of labels, while each of the remaining vertices
$u\in U$ is assigned the label set $\G(u) \subseteq V$. We thus have a \CP instance on $T$. If $u_1,u_2,\ldots,u_{|U'|}$ and
$V'$ is feasible for {\it Nested Neighborhoods} on $G$, then we can construct a feasible solution for \CP on $T$, with the
same cost, by choosing a permutation of $V$ that has the labels of $\G(u_{|U'|})\cap V'$ first, followed by those of
$\G(u_{|U'|-1})\cap V'\setminus \G(u_{|U'|})$ and so on. Thus the {\it Nested Neighborhoods} problem reduces to the \CP
problem on stars.

Conversely, if $T$ is star in an instance of \CP, with $\S$ as the label set of the non-leaf vertex $r$ and $\S_i$ as the
label set of each leaf $u_i$, then we construct a {\it Nested Neighborhoods} instance as follows. The bipartition has the
vertex sets $U$, which consists of all the leaf nodes of $T$, and $V$ which consists of the set of labels $\S$ on $r$.
A vertex $u_i \in U$ is connected by an edge to a vertex $v_s\in V$, if the corresponding label $s\in \S$ belongs to the
label set $\S_i$ of $u_i$. Using an argument similar to that in the previous paragraph, it can be shown that each
feasible solution to \CP on $T$ has a corresponding feasible solution to {\it Nested Neighborhoods} on $G$,
with the same cost. We thus have the following result.

\begin{thm}\label{equivalence}
The {\it Nested neighborhoods} problem is equivalent to the \CP problem on an appropriate star. 
\end{thm}

We note that these are approximation preserving reduction. From now on, we deal with the {\it Nested Neighborhoods} problem.

\section{Edge Bicliques Problem}
Let $G=(U,V,E)$ be a bipartite graph with $U$ and $V$ as its bipartition and $E$ as its set of edges. If $B$ is a subset of
the vertex set ($U\cup V$), the subgraph induced by $B$ is said to be a {\it biclique} if $uv \in E$ for all
$u \in B\cap U$ and $v \in B\cap V$.
The {\it Maximum Edge Biclique} ($EBCS$) problem asks for a subgraph of a given bipartite graph, which is a biclique and has
the largest number of edges.

\begin{lem}\label{lowerbound}
Let $G=(U,V,E)$ be a bipartite graph, and let $\E$ and $\O$ be the optimal values of the {\it EBCS} and
the \NN problem on $G$. Then $\E \leq \N$.
\end{lem} 
\begin{Proof}
Suppose that $U'=\{u_1,u_2,\ldots,u_k\}\subseteq U$ and $V'=\{v_1,v_2,\ldots,v_l\}\subseteq V$ is a biclique. Since
$\G(u_1)\cap V' = \G(u_2)\cap V' = \ldots = \G(u_k)\cap V'$, this corresponds to a feasible solution of the {\it Nested Neighborhoods}
problem, with the same cost. 
\end{Proof}

Note that the above proof shows the stronger result that every feasible solution to {\it EBCS} has a corresponding feasible
solution to \NN with at least as much cost. 

\begin{lem}\label{upperbound}
Let $G=(U,V,E)$ be a bipartite graph. If it has a feasible solution to \NN of cost $c$, then $G$ contains a biclique
with at least $\frac{c}{H_n}$ edges, where $H_n$ denotes the $n^{th}$ harmonic number and $|U|=n$.
\end{lem}
\begin{Proof}
Let $U'$ and $V'$ be a feasible solution to the \NN problem of cost $c$, with $\{u_1,u_2,\ldots,u_k\}=U'$ and such that
$\G(u_1)\cap V' \supseteq \G(u_2)\cap V' \supseteq \ldots \supseteq \G(u_k)\cap V'$. Each vertex subset of the form
$u_1,u_2,\ldots,u_i$ along with $V'\cap_{j=1}^{j=i}\G(u_i)$ forms a biclique. It is easy to see that if the largest
biclique in the subgraph $P$, induced by $U'\cup V'$, contains $u_i$, then it also contains all vertices $u_j$ for $j\leq i$.
Let $\e$ be the size of the largest biclique in $P$ and let $y_i$ denote $|\G(u_i)\cap V'|$. The biclique induced by
$u_1,u_2,\ldots,u_i$ and $V'\cap_{j=1}^{j=i}\G(u_i)$ has $i\times y_i$ edges. Hence, for each $i=1,\ldots,k$, $y_i \leq \e/i$. 
We now have
\begin{eqnarray*}
c & = & y_1 + y_2 + \ldots + y_k \\
   & \leq & (1 + \frac{1}{2} + \frac{1}{3} + \ldots + \frac{1}{k})\e \\
   & \leq & H_n \e.
\end{eqnarray*} 
\noindent This proves the lemma.
\end{Proof}

Combining lemma~(\ref{lowerbound}) and lemma~(\ref{upperbound}) we get the following.

\begin{eqnarray*}
\E  \leq & \N & \leq H_n \E
\end{eqnarray*}

There are graphs for which the inequality on the right is tight. Consider the bipartite graph $G=(U,V,E)$,
with $U=\{u_1,u_2,\ldots,u_n\}$ and $V=\{v_1,v_2,\ldots,v_n\}$ and the edges defined by the relation
$\G(u_i) = \{ v_1,v_2,\ldots,v_{\frac{n}{i}}\}$. It is easily seen that every edge occurs in the optimal solution
to \NN. Thus $\N = n+(n/2)+(n/3)+\ldots+(n/n) = nH_n$. Further, if $k$ is the largest index of 
a vertex in $U$ in an optimal solution to $EBCS$, then every vertex $u_i$ is in the optimal for $i\leq k$, so that
$\E = k(n/k) = n$. Thus $\frac{\N}{\E} = H_n$ for this graph.

\section{Hardness of Common Prefix on Stars}
\def\a{\alpha}

We will need the following result of Feige and Kogan.

\begin{thm}\label{feigekogan}{\sf Feige-Kogan~\cite{Feige}}\\
If the maximum edge biclique problem can be approximated within a factor of $2^{(\log n)^{\delta}}$ for every constant 
$\delta >0$, then 3-SAT can be solved in time $2^{n^{3/4+\epsilon}}$ for every constant $\epsilon >0$.
\end{thm}

Suppose that there is an algorithm that approximates \NN on stars within a factor of $\a$, i.e. if it returns the value $A$,
then $\O \leq A \leq \a\N$. Then using lemma-(\ref{upperbound}), we know that the bipartite graph contains a
feasible solution to $EBCS$ of size $A'$, such that $A\leq H_n A'$. We then get an $\a/H_n$ factor algorithm for
$EBCS$, since

\begin{eqnarray*}
A' & \geq & \frac{A}{H_n} \\
   & \geq & \frac{\a}{H_n}\N \\
   & \geq & \frac{\a}{H_n}\E.
\end{eqnarray*}

Thus, using theorems (\ref{equivalence}) and (\ref{feigekogan}), we get the following hardness result.

\begin{thm}
If the \CP  problem for stars can be approximated within a factor of $2^{(\log n)^{\delta}-\log\log n}$ for every constant 
$\delta >0$, then 3-SAT can be solved in time $2^{n^{3/4+\epsilon}}$ for every constant $\epsilon >0$.
\end{thm}

\section{Acknowledgments}
We thank Ravindra Guravannavar for posing this problem.

\end{document}